\documentclass{commatDV}

\usepackage{DLde}
\usepackage{graphicx}

\title{Action of vectorial Lie superalgebras on some split supermanifolds}

\author[Arkady Onishchik]{\fbox{Arkady Onishchik}}

\affiliation{N/A}

\abstract{The ``curved'' super Grassmannian is the supervariety of subsupervarieties of purely odd dimension $k$ in a~supervariety of purely odd dimension $n$, unlike the ``usual'' super Grassmannian which is the supervariety of linear subsuperspacies of purely odd dimension $k$ in a~superspace of purely odd dimension $n$. The Lie superalgebras of all and Hamiltonian vector fields on the superpoint are realized as Lie superalgebras of derivations of the structure sheaves of certain ``curved'' super Grassmannians.}



\VOLUME{30}
\NUMBER{3}
\firstpage{41}
\DOI{https://doi.org/10.46298/cm.10455}

\begin{paper}

\section*{Preface of the editor}\label{S:0}
The manuscript of this paper appeared as a preprint in proceedings of the ``Seminar on Supersymmetries'', see \url{http://staff.math.su.se/mleites/sos.html}, and in Russian, see [52] in the list of references in the jubilee paper \cite{AJ}. I updated the references; the ones I added are endowed with an asterisk. The abstract and  comments are due to me. For a~comprehensive description of simple Lie superalgebras of vector fields over algebraically closed fields of any characteristic, see~ \cite{BGLLS}. \textit{D. Leites}

\section{Introduction}

The Lie superalgebra $W_n:=\Der \Lambda _\Cee[\xi _1,\ldots ,\xi _n]$ consisting of all vector fields on the superpoint $\cC^{0,n}$ is isomorphic, as shown in~\cite{3},  to the Lie superalgebra of vector fields, i.e., the global derivations of the structure sheaf, see~\cite{2},  of a~split complex supermanifold $\mathcal{CG}r_{n-1}^n$ determined by the tautological vector bundle of rank $n-1$ on the complex Grassmann manifold $\text{Gr}_{n-1}^n$. 

The Lie superalgebra $H _n$, the subsuperalgebra of $W_n$ consisting of Hamiltonian vector fields on the superpoint $\cC^{0,n}$, is isomorphic to the Lie superalgebra of vector fields on a~split complex supermanifold $\mathcal{CQ}^{n-2}$  associated with a~vector bundle of rank $n-1$ orthogonal to the tautological bundle on the quadric $Q^{n-2}\subset \Cee P^{n-1}$.

However, the method used in~\cite{3}, \cite{4} does not allow one to indicate explicitly these isomorphisms. In this paper I explicitly construct the $W_n$- and $H _n$-actions on the supermanifolds $\mathcal{CG}r_{n-1}^n$ and $\mathcal{CQ}^{n-2}$; I give a~new version of the proof of the above results.

D.~Leites told me that the supermanifolds $\mathcal{CG}r_{n-1}^n$ and $\mathcal{CQ}^{n-2}$ are the simplest examples of  what Manin called ``curved'' Grassmannians and ``curved''  quadrics, see \cite{Ma}. They were introduced in~\cite{1}. (Compare with the Grassmannians of \textbf{linear} subsuperspaces in a linear superspace, see \cite[Ch.4, \S ~3]{Ma}. For the complete list of homogeneous superdomains associated with the known Lie superalgebras of polynomial growth, see \cite{L}. \textit{D.L.})

\section{Superization of a~construction due to Serre}\label{s1}

In this section, I superize a~construction Serre introduced in~\cite{6}. This enables us to interpret elements of a~Lie superalgebra as Hamiltonian vector fields on the superpoint.

Let $V$ be a~purely odd vector space over $\Cee$,   and $\overline V$ a~second copy of the same space considered as purely even. The change of parity $V\tto \overline V$ will be denoted by $x\mapsto \overline x$ on every non-zero $x\in V$. 

Construct the Koszul complex of the $\Zee$-graded algebra 
\[
{A:=S(\overline V \oplus V)=S(\overline V)\otimes \Lambda(V)}
\]
which can be naturally considered as a~free supercommutative superalgebra. There exists a~unique derivation $d\in \Der _{-1}A$ such that $dx=\overline x$ and $d \overline x=0$ for any $x\in V$. Obviously, $d^2=0$.

Consider the $Z$-graded Lie superalgebra $W(V)=\Der \Lambda(V)$. Any element $\delta \in W(V)$ can be uniquely extended to a~derivation $\tilde \delta \in \Der A$ such that $[\tilde \delta ,d]=0$, see \cite{5}. The correspondence $\delta \mapsto \tilde \delta $ is a~faithful linear representation of the Lie superalgebra $W(V)$ in $A$.

Let $\omega \in S^2(\overline V)$ be a~nondegenerate bilinear form. Set
\[
H(\omega ):=\{\delta \in W(V)\mid \tilde \delta (\omega )=0\}.
\]
Then $H(\omega )$ is a~$\Zee$-graded subalgebra in $W(V)$ called the  \textit{Lie superalgebra of Hamiltonian vector fields}. Set
\[
DH(\omega ):=\{\delta \in W(V)\mid \tilde \delta (\omega )=\varphi \omega \text{ for some }\varphi \in A\}.
\]
Clearly, $DH(\omega)$ is a~$\Zee$-graded subalgebra of $W(V)$, and $H(\omega)$ is its ideal.

Hereafter we assume that $\dim V=n$. Set $W_n:=W(V)$, ${H_n:=H(\omega)}$, $DH_n:=DH(\omega )$ since these algebras are determined, up to an isomorphism, by $\dim V$. In $V$, select a~basis $\xi _1,\ldots ,\xi _n$. Then the elements $x_i=\overline \xi _i=d \xi _i$, where $i=1,\ldots ,n$, constitute a~basis in $\overline V$ and
\[
A=\Cee[x_1,\ldots ,x_n]\otimes \Lambda[\xi _1,\ldots ,\xi _n].
\]
Obviously, $d=\sum\limits _{1 \leqslant i \leqslant n}x_i  \partial_{\xi _i}$. Notice that $ \partial_{\xi _i} \in \Der _{-1}A$ coincides with the extension $\tilde\partial_{\xi _i}$ of $\partial_{\xi _i} \in W(V)_{-1}$.

\ssbegin{Lemma}\label{l1.1} \textup{1)} If $\tilde \delta (\omega )=\varphi \omega $ for some $\delta \in W(V)$, $\varphi \in A$, then $\varphi \in \Cee$.

\textup{2)} There is the following semidirect sum decomposition:
\[
DH(\omega )=H(\omega )\ltimes\Cee E ,\quad \text{where}\quad E =\sum\limits _{1 \leqslant i \leqslant n}\xi _i  \partial_{\xi _i}.
\]
\end{Lemma}

\begin{proof} 1) We may assume that $(x_i)_{i=1}^n$ is a~basis in which $\omega $ is the form
\[
\omega =\sum\limits _{1 \leqslant i \leqslant n}x_i^2=\sum\limits _{1 \leqslant i \leqslant n}(d \xi _i)^2.
\]
Then we have
\[
\tilde \delta \omega =2 \sum\limits _{1 \leqslant i \leqslant n}x_i \tilde \delta x_i=(-1)^k2 \sum\limits _{1 \leqslant i \leqslant n}x_id(\delta \xi _i) \quad \text{for any}\quad \delta \in W(V)_k.
\]
Setting $h_i=\delta \xi _i$ we get
\[
\tilde \delta \omega =(-1)^k2 \sum\limits _{1 \leqslant i,j \leqslant n}x_ix_j \frac {\partial h_j}{\partial \xi _i},\quad \text{where}\quad \frac {\partial h_j}{\partial \xi _i} \in \Lambda ^k(V).
\]
If $\delta $ satisfies the conditions of Lemma~\ref{l1.1}, then $\varphi =(-1)^k2 \frac {\partial h_i}{\partial \xi _i}$ for any $i=1,\ldots ,k$. In particular, $\varphi \in \Lambda ^k(V)$. Further, for any $i$, we have $\frac {\partial \varphi }{\partial \xi _i}=(-1)^k2 \frac {\partial ^2h_i}{\partial \xi _i^2}=0$. Therefore, $\varphi \in \Cee$.

2) Observe that $E (\omega )=2 \omega $, and so $E \in D H(\omega )_0$. Further, if $\delta \in DH(\omega )$, then, due to the proved above, $\tilde \delta \omega =c \omega $, where ${c\in \Cee}$. Hence, $\left (\tilde \delta -\frac 12\,c  E \right )(\omega )=0$, i.e., $\delta =\delta _0+\frac 12\,c E $, where ${\delta _0\in H(\omega )}$.
\end{proof}

\section{Vector bundles over $\Cee P^{n-1}$ and $Q^{n-2}$ and supermanifolds}\label{s2}

Define some special supermanifolds associated with vector bundles over $Q^{n-2}\subset \Cee P^{n-1}$ and  $\Cee P^{n-1}$.

Let $\dim V=n$, and $P(V)$ the corresponding protective space. Let us assume that the nonzero elements of $V^*$ are odd and those of $\overline V{}^*$ are even. In $V$, select a~basis $e_1,\ldots ,e_n$, and consider the dual bases $\xi _1,\ldots ,\xi _n$ and $x_1,\ldots , x_n$ of $V^*$ and $\overline V{}^*$, respectively. In the notation of~\S~\ref{s1} (applied to $V^*$ and $\overline V^*$) we have $x_i=\overline \xi _i$. 

The elements $x_1,\ldots ,x_n$ are homogeneous coordinates on $P(V)$; this means that the stalk $\cF_z^a$ of the structure sheaf $\cF^a$ of the algebraic variety $P(V)$ at $z\in P(V)$ is a~subring of the field $\Cee(V)=\Cee(x_1,\ldots ,x_n)$ consisting of elements of the form $f/g$, where $f,g$ are homogeneous polynomials of the same degree in $\Cee[x_1, \ldots ,x_n] = S(\overline V^*)$ and $g(z) \ne 0$. 

Consider the trivial vector bundle $P(V) \times V^*$ and its subbundle $E \subset P(V)\times V^*$ consisting of the pairs $(\Cee x,y)$, where $x\in V \setminus \{0\}$ and $y\in \Ann x=\{\alpha \in V^*\mid \alpha (x)=0\}$. Clearly, $E$ is an algebraic vector bundle of rank $n-1$ over $P(V)$ with the fiber 
\[
E_{\Cee x}=\Ann x, \text{~~ where $x\in V \setminus \{0\})$.}
\] 

The map $\Cee x\mapsto \Ann x$ identifies $P(V)$ with the Grassmann variety $\text{Gr}_{n-1}(V^*)$, which consists of $(n-1)$-dimensional subspaces in $V^*$, and $E$ is the tautological bundle over this Grassmann variety. 

Let $\cE^a \subset \cF^a \otimes V^*$ be the sheaf of germs of the polynomial sections of $E$. The variety $P(V)$ can be endowed with two structures of a~split complex algebraic supervariety: one is  determined by the sheaf ${\hat \cO{}^a=\cF^a \otimes \Lambda(V^*)}$, and the other by its subsheaf $\cO^a=\Lambda(\cE^a)$. 

Besides, $\hat \cO=\cF \otimes \Lambda(V^*)$ and $\cO=\Lambda (\cE) \subset \hat \cO$, where $\cF$ is the sheaf of holomorphic functions on $P(V)$ and $\cE$ is the sheaf of germs of holomorphic sections of $E$, determine two structures of a~split complex analytic supermanifold associated with $P(V)$.

Consider the superalgebra
\[
A:=S(\overline V{}^*)\otimes \Lambda(V^*)=\Cee[x_1,\ldots ,x_n]\otimes \Lambda[\xi _1,\ldots ,\xi _n]
\]
and the derivation $d\in \Der _{-1}A$ constructed in~\S~\ref{s1} (with $V$ replaced by $V^*$ in these constructions). Let
\[
B:=\Cee(V)\otimes \Lambda(V^*)=\Cee(x_1,\ldots ,x_n)\otimes \Lambda[\xi _1,\ldots ,\xi _n]
\]
be the localization of $A$ with respect to the multiplicative system $S(\overline V{}^*)\setminus \{0\}$. The algebra $B$ has a~natural supercommutative superalgebra structure, and $d$ can be uniquely extended to an odd derivation of $B$, which we will denote also by $d$. Clearly, 
\[
\hat \cO{}_z^a=\cF_z^a \otimes \Lambda(V^*)\subset B\text{~~ for any $z\in P(V)$.}
\]

\ssbegin{Lemma}\label{l2.1} We have $\cO_z^a=\hat \cO{}_z^a \cap \Ker D$ for any $z\in P(V)$.
\end{Lemma}

\begin{proof} For any $x\in V \setminus \{0\}$, consider the derivation 
\[
d_x=\sum\limits _{1 \leqslant i \leqslant n}x_i\partial_{\xi _i} \in \Der _{-1}\Lambda(V^*)
\]
 uniquely determined by the condition $d_x(\xi )=\xi (x)$ for any $\xi \in V^*$. Clearly, 
\[
\Ker d_x=\Lambda(\Ann x)=\Lambda (E_{\Cee x}). 
\]
Let $u:=\sum\limits _{1 \leqslant i \leqslant r}\varphi _iv_i\in B$, where $\varphi _i\in \cF^a$ and $v_i\in \Lambda(V^*)$. Considering $u$ as a~function on $V \setminus \{0\}$ with values in $\Lambda(V^*)$ we see that ${u(x)=\sum\limits _{1 \leqslant i \leqslant r}\varphi _i(\Cee x)v_i}$ at any point $x$ from its domain implying
\[
d_xu(x)=\sum\limits _{1 \leqslant i \leqslant r}\varphi _i(\Cee x)d_xv_i=\sum\limits _{1 \leqslant i \leqslant r}\varphi _i(\Cee x)\sum\limits _{1 \leqslant j \leqslant n}x_j \frac {\partial v_i}{\partial \xi _j}.
\]
On the other hand,
\[
du=\sum\limits _{1 \leqslant i \leqslant r}\varphi _idv_i=\sum\limits \varphi _i \sum\limits _{1 \leqslant i \leqslant n}x_j \frac {\partial v_i}{\partial \xi _j}.
\]

Therefore, $(du) (x) = d_x u(x)$ on a~Zariski open subset of $V$. Hence, $du=0$ if and only if  $u(x) \in \Lambda(E_{\Cee x})$ for all $x\in V$ from the domain of~$u$.
\end{proof}

Now, define a~subsupermanifold of $(P(V),\cO^a)$ whose underlying manifold is the quadric $Q \subset P(V)$. Let $\omega $ be a~non-degenerate quadratic function on $V$; it can be considered as an element of $S^2(\overline V{}^*)$. Let $Q$ be the quadric in $P(V)$ given by the equation $\omega =0$ and $E'=E|_Q$ the restriction of $E$ to $Q$. If we identify $V^*$ with $V$ with the help of the non-degenerate symmetric bilinear form corresponding to $\omega$, then $E'$ is identified with the subbundle of $Q \times V$ orthogonal to the tautological line bundle over $Q$.

Let $\cF'^a$ and $\cE'^a$ (resp. $\cF'$ and $\cE'$) be the sheaves of polynomial (resp. holomorphic) functions on $Q$ and polynomial (resp. holomorphic) sections of $\cE'$, respectively. Set ${\cO'^a:=\Lambda (\cE'^a})$ and $\cO':=\Lambda (\cE')$.

Let us give a~description of $\cO'^a$ similar to the above description of $\cO^a$. For this, consider the superalgebra
\[
A' :=A/\omega A=\Cee[\hat Q]\otimes \Lambda(V^*),
\]
where $\Cee[\hat Q]:=S(\overline V{}^*)/\omega S(\overline V{}^*)$ is the algebra of polynomial functions on the cone $\hat Q \subset V$ given by the equation $\omega =0$;   the localization of $A'$ is ${B' :=\Cee(\hat Q)\otimes \Lambda(V^*)}$. The algebras $\cF{}'^a_z$ and $\cO_z'^a$, where $z\in Q$, are embedded into $B'$. The derivation $d$ transforms $\omega A$ into itself, and therefore determines an odd derivation $d'$ of $A'$ and $B'$. Lemma~\ref{l2.1} implies that
\[
\cO_z'^a=(\cF_z'^a \otimes \Lambda(V^*))\cap \Ker d' \quad \text{for any}\quad z\in Q.
\]

\section{Several remarks on vector fields on algebraic and analytic supervarieties}\label{s3}

Let $M$ be a~nonsingular complex algebraic variety, $E$ an algebraic vector bundle over $M$. Denote by $\cF^a$, $\cT^a$, $\cE^a$ the structure sheaf on $M$, the tangent sheaf on $M$, and the locally free algebraic sheaf corresponding to $E$, respectively. We denote by the same letters without the superscript $a$ the corresponding analytic sheaves on $M$.

In particular, $(M,\cF)$ is the complex analytic manifold corresponding to the algebraic variety $M$. The sheaves $\cO^a=\Lambda _{\cF^a}(\cE^a)$ and $\cO=\Lambda _\cF(\cE)$ rig $M$ with structures of a split algebraic and a split analytic supervariety, respectively. We call the sheaves of $\Zee$-graded Lie superalgebras $\Der \cO^a$ and $\Der \cO$  the \textit{sheaves of vector fields} on these supervarieties.

\ssbegin{Lemma}\label{l3.1} There exists a~natural injective homomorphism of shea\-ves of $\Zee$-graded Lie superalgabras $\Der \cO^a\tto \Der \cO$.
\end{Lemma}

\begin{proof}
As shown in~\cite{2}, for any $k\in \Zee$, every $\gamma \in \Der _k\cO$ can be identified  with a~pair $(\gamma _0,\gamma _1)$, where 
\be\label{pair}
\text{$\gamma _0\in \Hom _\cF(\cE,\Lambda ^{k+1}(\cE))$ and $\gamma _1\in \Hom _\Cee(\cF,\Lambda ^k(\cE))=\cT \mskip\medmuskip{\otimes}_\cF\mskip\medmuskip \Lambda ^k(\cE)$}
\ee  
such that
\begin{equation}
\label{1}
\begin{array}{l}
\gamma _0(\varphi s)=\gamma _1(\varphi )s+\varphi \gamma _0(s)\\
\gamma _1(\varphi \psi )=\gamma _1(\varphi )\psi +\varphi \gamma _1(\psi )
\end{array}
\qquad \text{for any }s\in \cE \text{ and }\varphi ,\psi \in \cF.
\end{equation}

A similar statement holds also for $\Der \cO^a$. 

Since the sections of the sheaves 
\[
\text{$\Hom _{\cF^a}(\cE^a,\Lambda ^{k+1}(\cE^a))$ (resp. $\Hom _\cF(\cE,\Lambda ^{k+1}(\cE))$)}
\]
 are algebraic (resp. holomorphic) homomorphisms of vector bundles ${E\tto \Lambda ^{k+1}(E)}$, there is a~natural embedding
\[
\Hom _{\cF^a}(\cE^a,\Lambda ^{k+1}(\cE^a)) \tto \Hom _\cF(\cE,\Lambda ^{k+1}(\cE)).
\]
Further, the natural embedding $\cT^a\tto \cT$ induces the embedding
\[
\cT^a \mskip\medmuskip {\otimes }_{\cF^a}\mskip\medmuskip \Lambda ^k(\cE^a)\tto \cT \mskip\medmuskip {\otimes }_\cF \mskip\medmuskip \Lambda ^k(\cE).
\]
Therefore, to every pair $(\gamma _0,\gamma _1) \in \Hom _{\cF^a}(\cE^a,\Lambda ^{k+1}(\cE^a)) \times \Hom _\Cee(\cF^a,\Lambda ^k(\cE^a))$ that satisfies conditions similar to~\eqref{1}, there corresponds a~pair,  see \eqref{pair},  satisfying~\eqref{1}, as is easy to verify. 

This is the desired embedding ${\Der \cO^a\tto \Der \cO}$.
\end{proof}

\sssbegin{Corollary} There exists an injective homomorphism of $\Zee$-grad\-ed Lie superalgebras of vector fields $\fd^a\tto \fd$, where $\fd^a=\Gamma (M,\Der \cO^a)$ and $\fd=\Gamma (M,\Der \cO)$.
\end{Corollary}

The results of~\cite{6} imply that the homomorphism $\fd^a\tto \fd$ is an isomorphism for any projective variety $M$.

\section{$W_n$ and $DH_n$ as vectorial Lie superalgebras}

Now we are able to determine the action of $W_n$ and $DH_n$ on the supermanifolds constructed in~\S~\ref{s2}. Retaining the notation of~\S~\ref{s2} let us first prove the following statement.

\ssbegin{Lemma}\label{l4.1} Let $\gamma \in \Der B$ be such that $\gamma V \subset \Lambda(V)$, so that
\[
\text{${\gamma x_i=\sum\limits _{1 \leqslant i \leqslant n}v_{ij}x_j}$, where $i=1,\ldots ,n$ and $v_{ij} \in \Lambda(V)$.}
\]
 Then, $\gamma (\hat \cO{}_z^a)\subset \hat \cO{}_z^a$ for all $z\in P(V)$.
\end{Lemma}

\begin{proof} Let $u=\varphi v$, where $\varphi \in \cF_z^a$ and $v\in \Lambda(V)$. Then, \[
\gamma (u)=\gamma (\varphi )v+\varphi \gamma (v). 
\]
Therefore, it suffices to verify that $\gamma (\varphi ) \in \hat {\cO}{}_z^a$. Let us express $\varphi $ in the form $f/g$, where $f,g\in \Cee[x_1,\ldots ,x_n]$ are homogeneous polynomials of the same degree and $g(z) \ne 0$. Our statement follows easily from  the hypothesis on $\gamma $ and the identity 
\[
\gamma (\varphi )=\frac 1{g^2}\,(g \gamma (f)-f \gamma (g)).
\]
Now, let $\delta \in W_n=W(V^*)$. As we have seen in~\S~\ref{s1}, $\delta $ can be uniquely extended to a~derivation $\tilde \delta $ of $A=S(\overline V{}^*)\otimes \Lambda(V^*)$ such that $[\tilde \delta ,d]=0$. Obviously, the action of $\tilde \delta $ can be uniquely extended to the localization $B$ of $A$, so (see Proof of Lemma~\ref{l1.1})
\[
\tilde \delta x_i=\tilde \delta d \xi _i=d \delta \xi _i=\pm \sum\limits _{1 \leqslant j \leqslant n}x_j \frac {\partial h_i}{\partial \xi _j},\quad \text{where}\quad h_i=\delta \xi _i\in \Lambda(V).
\]
By Lemma~\ref{l4.1} $\tilde \delta $ transforms all the algebras $\hat \cO{}_z^a$, where $z\in P(V)$, into themselves. Since $\tilde \delta $ transforms $\Ker d$ into itself, then Lemma~\ref{l2.1} implies that $\delta (\cO_z^a)\subset \cO_z^a$ for all $z\in P(V)$. Therefore, $\delta $ determines a~global derivation of $\cO^a$. We have constructed a~map 
\[
W_n\tto \fd^a=\Gamma (P(V),\Der \cO^a). 
\]
As is easy to see, theis map is an injective homomorphism of $\Zee$-graded Lie superalgebras. By Lemma~\ref{l3.1} we also have an injective homomorphism $W_n\tto \fd=\Gamma (P(V),\Der \cO)$.
\end{proof}

We similarly construct an injective homomorphism 
\[
DH_n\tto \fd'=\Gamma (Q,\Der \cO'). 
\]
If ${\delta \in DH_n}$, then $\tilde \delta \in \Der A$ transforms $\omega A$ into itself, and therefore determines a~derivation of $A'$ from~\S~\ref{s2} which extends to $B'$ and yields a~derivation of $\cO'^a$.

\ssbegin{Theorem}\label{t4.2} The above-constructed homomorphisms $W_n\tto \fd$ and $DH_n\tto \fd'$ are isomorphisms if $n \geqslant 2$ and $n \geqslant 5$, respectively.
\end{Theorem}

\begin{proof} By the proved above we may assume that the finite-dimensional $\Zee$-graded Lie superalgebras $\fd$ and $\fd'$ contain $W_n$ and $DH_n$, respectively, as subalgebras. Therefore, it suffices to prove that $\fd$ and $\fd'$ are transitive and irreducible and $\fd_k=(W_n)_k$ and $\fd'_k=(DH_n)_k$ for $k= -1, 0$ (see~\cite[Theorem 4]{5}). A proof of these statements is contained in~\cite{3}, \cite{4}. This proof essentially depends on Lemma 4.1.1 in~\cite{2} and actually reduces to the calculation of $\fd_0$ and $\fd_0'$ and also $\fd_0$-module $\fd_{-1}$ and $\fd_0'$-module $\fd_{-1}'$ with the help of Bott's theorem.
\end{proof}

\end{paper}
\begin{references}

\bibitem[AVGDZKLST*]{AJ}
D. N. Akhiezer, \`E. B. Vinberg, V. V. Gorbatsevich, V. G. Durnev, R. Zulanke, L. S. Kazarin, D. A. Leites, V. V. Serganova, V. M. Tikhomirov,  Arkadii L'vovich Onishchik (on his 70th birthday). Russian Math. Surveys, 58:6 (2003), 1245--1253.

\bibitem[BGLLS*]{BGLLS}
Bouarroudj S., Grozman P., Lebedev A., Leites D., Shchepochkina I., Simple vectorial Lie algebras in characteristic $2$ and their superizations. Symmetry, Integrability and Geometry: Methods and Applications
(SIGMA),   16 (2020), 089, 101 pages; \texttt{arXiv:1510.07255}.

\bibitem[K]{5}
Kac V. G., Lie superalgebras. Adv. Math. 1977. \textbf{26}, No. 1, 8--96.

\bibitem[KL*]{1}
Kirillova R.~Yu., Leites D.~A., Instantons with gauge supergroup. In: Problems of
nuclear physics and cosmic rays, 1985, no.~24 (D.~V.~Volkov
Festschrift), Kharkov Univ. Press, Kharkov, 34--40 (in Russian) MR 88h:53032.

\bibitem[L*]{L}
Leites D.~A., On unconventional integration on supermanifolds and cross ratio
on classical superspaces. In: E.~Ivanov, S.~Krivonos, J.~Lukierski,
A.~Pashnev (eds.) Proceedings of the International Workshop
``Supersymmetries and Quantum Symmetries", September 21--25, 2001,
Karpacz, Poland. JINR, Dubna, 2002, 251--262;
\texttt{arXiv:math.RT/0202194}.

\bibitem[Ma*]{Ma}
Manin Yu., {\it Gauge field theory and complex geometry}. Second
edition. Springer-Verlag, Berlin, 1997. xii+346 pp.

\bibitem[O]{2}
Onischik A.~L., Transitive Lie superalgebras of vector fields. Communications in Mathematics \textbf{30} (2022), no. 3, 25--40.

\bibitem[OS1*]{OS1}
 Onishchik A.~L., Serov A. A., Lie superalgebras of vector fields on splittable flag supermanifolds, Dokl. Akad. Nauk SSSR 300 (1988), no. 2, 284-287 (in Russian); English translation: Soviet Math. Dokl. 37 (1988), no.3, 652--655. MR 89i:32067.

\bibitem[OS2*]{OS2}
Onishchik, A. L.; Serov, A. A. Holomorphic vector fields on super-Grassmannians. In: Lie groups, their discrete subgroups, and invariant theory. Adv. Soviet Math., 8, Amer. Math. Soc., Providence, RI, 1992, 113--129.

\bibitem[S1]{3}
Serov A. A., Vector fields on split supermanifolds, 22--81. In: Leites D. (ed.) ``Seminar on Supersymmetries'', 20/1987-26, see \url{http://staff.math.su.se/mleites/sos.html}. For a~summary, see  \cite{OS1}, \cite{OS2}.

\bibitem[S2]{4}
Serov A.~A., A realization of the Hamiltonian Lie superalgebra. In: Onishchik A. (ed.) Problems in group theory and homological algebra, 164–167, Matematika, Yaroslav. Gos. Univ., Yaroslavl, 164--167 (in Russian) MR1174807.


\bibitem[Se]{6}
Serre J.-P., G\'eometrie alg\'ebraique et g\'eometrie analytique. Ann. Inst. Fourier. 1956. \textbf{6},  1--42.


\end{references}
